\documentclass{article}

\usepackage[T1]{fontenc}
\usepackage[latin9]{inputenc}
\usepackage{geometry}
\usepackage{float}
\usepackage{graphicx}

\makeatletter

\floatstyle{ruled}
\newfloat{algorithm}{tbp}{loa}
\providecommand{\algorithmname}{Algorithm}
\floatname{algorithm}{\protect\algorithmname}

\newenvironment{lyxcode}
{\par\begin{list}{}{
\setlength{\rightmargin}{\leftmargin}
\setlength{\listparindent}{0pt}% needed for AMS classes
\raggedright
\setlength{\itemsep}{0pt}
\setlength{\parsep}{0pt}
\normalfont\ttfamily}%
 \item[]}
{\end{list}}
\providecommand*{\code}[1]{\texttt{#1}}

\makeatother

\begin{document}

\title{Confederated Modular Differential Equation APIs for Accelerated Algorithm
Development and Benchmarking}

\author{Christopher Rackauckas\thanks{Department of Mathematics, University of California, Irvine, Irvine,
CA 92697, USA and Center for Complex Biological Systems, University
of California, Irvine, Irvine, CA 92697, USA (contact@chrisrackauckas.com)} ~and Qing Nie\thanks{Department of Mathematics, University of California, Irvine, Irvine,
CA 92697, USA and Center for Complex Biological Systems, University
of California, Irvine, Irvine, CA 92697, USA and Department of Developmental
and Cell Biology, University of California, Irvine, Irvine, CA 92697,
USA (qnie@math.uci.edu)}}
\maketitle
\begin{abstract}
Performant numerical solving of differential equations is required
for large-scale scientific modeling. In this manuscript we focus on
two questions: (1) how can researchers empirically verify theoretical
advances and consistently compare methods in production software settings
and (2) how can users (scientific domain experts) keep up with the
state-of-the-art methods to select those which are most appropriate?
Here we describe how the confederated modular API of DifferentialEquations.jl
addresses these concerns. We detail the package-free API which allows
numerical methods researchers to readily utilize and benchmark any
compatible method directly in full-scale scientific applications.
In addition, we describe how the complexity of the method choices
is abstracted via a polyalgorithm. We show how scientific tooling
built on top of DifferentialEquations.jl, such as packages for dynamical
systems quantification and quantum optics simulation, both benefit
from this structure and provide themselves as convenient benchmarking
tools.
\end{abstract}

\section{Introduction}

Differential equations are a ubiquitous modeling tool across the biological
to physical sciences. In most occasions, these equations have no analytical
solution and thus must be solved numerically. However, each of the
many families of integration methods utilize different structures
of the underlying equations to achieve differing problem-specific
efficiencies. Since performant handling of these equations is required
for large-scale scientific modeling, an increasingly pressing question
is two-fold: (1) how can researchers empirically verify theoretical
advances and consistently compare many different methods in production-software
settings and (2) how can users (scientific domain experts) keep up
with the state-of-the-art methods to select the most appropriate integrator
for their problem?

In this manuscript we will describe the confederated modular API of
the DifferentialEquations.jl \cite{christopher_rackauckas_differentialequations.jl_2017}
metapackage and show how it accelerates scientific progress by directly
addressing these two questions. We will begin by describing how the
modular package-generic API with automatic composability through multiple
dispatch mechanisms allows numerical methods researchers to readily
substitute any compatible method directly into full-scale scientific
applications for practical use, and also for testing and benchmarking.
We will discuss how the needs of three distinct groups of users, methods
developers, scientific package developers, and domain modelers, led
to the evolution of this mechanism and the problems it has solved.

On the other hand, the common API has led to a rapid expansion in
the available algorithm choices and this complexity has led to new
problems for end-users who tend to be less familiar with all of the
nuances of the field. To handle the growing complexity, we show how
the method choices have been abstracted by automatic algorithm specialization
via a type-hierarchy classification of differential equation structure
and an associated polyalgorithm. We show how the software ecosystem
built around this tooling, such as packages for quantum optics and
real-time robotics simulation, both benefit from this structure and
provide themselves as convenient benchmarking tools. We end by discussing
how further automation is being added to the framework and how this
software architecture can be replicated in other scientific domains.

\section{The Confederated Modular API}

\begin{figure}
\begin{centering}
\includegraphics[scale=0.5]{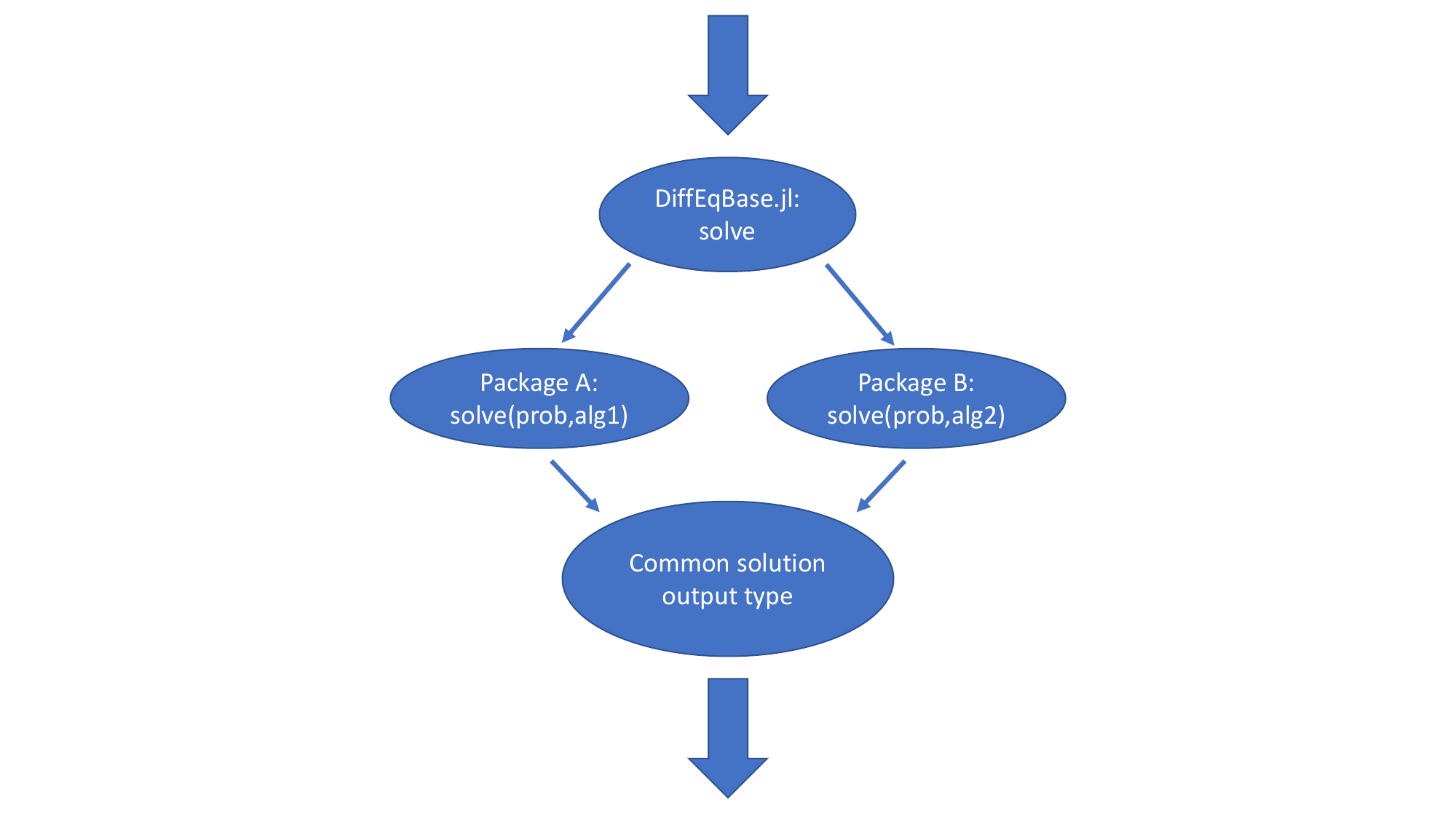}
\par\end{centering}
\caption{\textbf{The Confederated Modular API}. The confederated modular API,
known as the common API, is called by the user with the solve command.
This function is defined in DiffEqBase, but problem and algorithm
types are used to dispatch to separate packages. These packages interpret
the differential equation problem type (ODE, SDE, etc.) and solve
it using the documented algorithm. Each of these packages return a
common solution type with a defined API. The user then continues their
work using this common output. Note that if not algorithm is given
by the user, the algorithm for dispatch is decided automatically using
the algorithm described in Section \ref{sec:Polyalgorithms-for-Automatic}.
\label{fig:api}}

\end{figure}

The conmmon API is the API across all of the different solver packages
in DifferentialEquations.jl and is diagramed in Figure \ref{fig:api}.
DifferenitalEquations.jl's structure is defined in the base package
DiffEqBase.jl. Its main purpose is to hold the function definitions
which are utilized at all levels of the API. The Julia programming
language is built around a multiple dispatch system which allows for
specializing function calls via methods which are dependent on the
inputs \cite{bezanson_julia:_2017}. The multiple dispatch mechanism
is usually discussed in terms of its ability to allow for code specialization
for high performance. For example, the function call \code{+(2,5)}
lowers to a different method than \code{+(2.0,5.0)} which allows
a Julia programmer and the compiler to generate efficient LLVM byte
code that is efficient for each case. The Julia compiler will utilize
known type information at compilation time (while making use of type
inference) to allow for this structure to have no runtime overhead
which leads to performance comparable to statically-typed languages
like C or Fortran.

However, the confederated modular API utilizes the multiple dispatch
mechanism as a means for allowing input arguments to choose underlying
integration methods. The core function is the \code{solve} function.
This is defined as a \textquotedbl{}stub function\textquotedbl{} in
DiffEqBase.jl, i.e. a function without a method which thus has no
functionality. The documented API is that \code{solve} is called
via \code{solve(prob,alg)} where \code{prob} is some problem type
and \code{alg} is the algorithm type. For example, the \code{ODEProblem}
type has fields for \code{f}, \code{u0}, \code{tspan}, and \code{p}
which define the ODE system:

\[
u'=f(u,p,t),\thinspace\thinspace\thinspace u(t_{0})=u0,
\]

while the algorithm type is used simply to dispatch to the solver
algorithm to return a solution. An example of user code for solving
the Lorenz ODE system by the 5th order explicit Runge-Kutta method
\cite{tsitouras_runge-kutta_2011} is shown in Algorithm \ref{alg:lorenz}.

\begin{algorithm}
\begin{lyxcode}
function~lorenz(du,u,p,t)~~~

~~du{[}1{]}~=~10.0{*}(u{[}2{]}-u{[}1{]})~~~

~~du{[}2{]}~=~u{[}1{]}{*}(28.0-u{[}3{]})~-~u{[}2{]}~~~

~~du{[}3{]}~=~u{[}1{]}{*}u{[}2{]}~-~(8/3){*}u{[}3{]}~

end~

u0~=~{[}1.0;0.0;0.0{]};~tspan~=~(0.0,100.0)~

prob~=~ODEProblem(lorenz,u0,tspan)~

sol~=~solve(prob,Tsit5())
\end{lyxcode}
\caption{\textbf{Lorenz Equation Example. }Example user code solving Lorenz
ordinary differential equation using DifferentialEquations.jl. The
algorithm choice is given by the argument \protect\code{Tsit5()}
which causes \protect\code{solve} to dispatch to the solver defined
in the OrdinaryDiffEq.jl package. \label{alg:lorenz}}
\end{algorithm}

The key feature of this design is that any Julia code can extend the
solve function by defining a method dispatch on their algorithm type,
allowing for their code to be used to generate the solution. An example
of defining the Euler method for an ODEProblem is shown in Algorithm
\ref{alg:euler}.
\begin{lyxcode}
\begin{algorithm}
\begin{lyxcode}
struct~Euler~<:~DEAlgorithm~end

function~DiffEqBase.solve(prob::ODEProblem,alg::Euler,args...;

~~~~~~~~~~~~~~~~~dt~=~error(\textquotedbl{}dt~required~for~Euler\textquotedbl{}))

~~tspan~=~prob.tspan;~p~=~prob.p

~~n~=~Int((tspan{[}2{]}~-~tspan{[}1{]})/dt)~+~1

~~u~=~{[}prob.u0~for~i~in~1:n{]}

~~t~=~{[}tspan{[}1{]}~+~i{*}dt~for~i~in~0:n-1{]}

~~for~i~in~2:n

~~~~~~uprev~=~u{[}i-1{]}

~~~~~~tprev~=~t{[}i-1{]}

~~~~~~u{[}i{]}~=~uprev~+~dt{*}prob.f(uprev,p,tprev)

~~end

~~build\_solution(prob,alg,t,u)

end
\end{lyxcode}
\caption{\textbf{Dispatch Example.} An example code which defines a dispatch
for the \protect\code{solve} command and utilizes the Euler method
to solve the given ODE problem. \label{alg:euler}}

\end{algorithm}
\end{lyxcode}
This design means that the choices of methods available to the user
are not built into the function itself. Rather, every differential
equation solver method is defined through a common extension system,
and a large set of packages defining integrator methods comprises
the JuliaDiffEq organization. However, this extension system does
not have a core central authority as any author in any package can
extend the function like is shown in Algorithm \ref{alg:euler}. This
is why we describe this system as a confederated modular API (it is
colloquially referred to as the common API). Additionally, the common
API has documented options (abstol, reltol, etc.) and event-handling
choices to allow for all algorithms to act the same on the user's
input and thus abstract the package-free nature away from the user.

\section{Direct Benefits of the API Design}

This feature has greatly accelerated integrator methods development
in the Julia differential equations community for a few reasons. One
major reason is that it allows researchers to develop implementations
which are not part of a central repository. Academic success requires
that individuals build a portfolio of publications and, more recently,
software. In the highly competitive academic atmosphere, many individuals
simply cannot justify donating time to larger open-source projects.
The confederated modular API allows these individuals to build methods
in their own repositories under their own name, yet be utilized by
end users just like any other JuliaDiffEq solver. One prominent case
of this is LSODA.jl \cite{veltz_lsoda.jl_2017} which wraps a thread-safe
C++ re-write of the classic LSODA algorithm \cite{petzold_automatic_nodate,forst_liblsoda_2016},
but is not encompassed in the JuliaDiffEq organization since its utilization
and maintenance is tracked as part of the individual's academic achievements.

Another reason the confederated API has been a success is that it
allows for solver methods to be developed in isolation. This means
that a user can develop a useful addition to the available solver
methods without having to understand the organizational policies of
a larger open-source project. For example, DASSL.jl \cite{biernat_dassl.jl_2016}
is a re-write of the classic DASSL algorithm \cite{petzold_description_nodate}
into Julia which doesn't depend on any JuliaDiffEq resources other
than DiffEqBase.jl to define the extension. This has been successful
in recruiting earlier-stage academics, such as undergraduates in the
Google Summer of Code program, to be method contributors since all
that is required is the ability to understand the mathematical method
and write Julia code. In addition, since changing the solver package
choice only requires a change of input type, this allows for package-free
methods development tooling, such as convergence testing and benchmarking,
to be developed and systematically applied across the whole range
of dispatch choices (detailed in a later section).

Lastly, a recent addition to the JuliaDiffEq organization are the
diffeqr and diffeqpy packages for R and Python respectively that give
multi-language access to the common API. Thus any developer who implements
their differential equation solver method as a dispatch on the solve
common automatically gets access to scientific users in the Jupyter
(Julia, Python, and R) languages. This increases the possible scientific
impact of each individual's work, allowing for accelerated scientific
discovery as both users and methods developers do not need to worry
about developing cross-language interoperability for their method
to be impactful to the greater scientific community.

However, the confederation of solver packages has not just been advantageous
to methods developers. Downstream scientific modelers and their domain-specific
packages have benefitted since allowing the user to choose methods
then allows for user choices of packages without any extra support
required. An example implementation of a function which does parameter
optimization (via an $L^{2}$ loss function of the ODE timeseries
against a dataset) which works with any common interface ordinary
differential equation, delay differential equation (DDE), differential-algebraic
equation (DAE), or hybrid differential equation (including event handling)
is shown in Algorithm \ref{alg:l2loss}. We want to stress that this
compatibility with a wide array of packages and differential equation
types is not directly coded into our parameter estimation function.
Rather, this functionality comes just by allowing the user to give
the input problem type and the solver algorithm type and then all
package (and sometimes language) interop is handled in the solve dispatches.
\begin{lyxcode}
\begin{algorithm}
\begin{lyxcode}
function~parameter\_l2loss(prob,alg,t,data;kwargs...)

~~cost\_function~=~function~(p)~\#~Cost~function~to~return

~~~~tmp\_prob~=~remake(prob,p=p)~\#~Same~problem,~new~parameters

~~~~sol~=~solve(tmp\_prob,alg;saveat=t,save\_everystep=false,

~~~~~~~~~~~~~~~~~~~~~~~~~~~~~dense=false,kwargs...)~\#~Get~solution~at~t's

~~~~sum(abs2,data~-~sol)~\#~Get~L2~loss

~~end

end
\end{lyxcode}
\caption{\textbf{Generic $L^{2}$ Loss Function Generation.} This is an example
of a package code for parameter optimization. The function \protect\code{parameter\_l2loss}
returns a function that internally evaluates the ODE with given parameters
and calculates the $L^{2}$ loss against a data set. The returned
function can be given to a nonlinear optimization package to find
the optimal parameters. This function is agnostic to the input problem
and algorithm choice: \protect\code{prob} is the problem to solve
(\protect\code{ODEProblem}, \protect\code{DAEProblem}, \protect\code{DDEProblem},
etc.) and \protect\code{alg} can be an algorithm provided by any
package which has a dispatch for \protect\code{solve} on the given
problem type. \label{alg:l2loss}}
\end{algorithm}
\end{lyxcode}
This extensibility by allowing type inputs has been utilized in downstream
domain-modeling packages. For example, for solving the quantum dynamical
equations such as the Schrodinger partial differential equations (PDEs)
in QuantumOptics.jl \cite{kramer_quantumoptics.jl:_2018}, for systems
with smaller numbers of states one may want to make use of the implicit
BDF method available from the SUNDIALS C++ package \cite{alan_c._hindmarsh_sundials:_2005}
and wrapped via Sundials.jl, while to conserve memory for larger PDEs
one may want to resort to the low-storage Runge-Kutta implementations
in OrdinaryDiffEq.jl. QuantumOptics.jl allows the user to pass in
any common interface algorithm type, and thus the user can change
integrators (and thus the underlying solver package) by changing this
input value, allowing QuantumOptics.jl to be package-generic without
having to directly support any of the possible choices. In addition,
this allows domain-specific packages to add new methods as required
to accelerate problems in their domain. For example, DynamicalSystems.jl
\cite{datseris_dynamicalsystems.jl:_2018} is a package downstream
of DifferentialEquations.jl which utilizes the discrete evolution
problems (functional iteration), but added their own algorithm type
and dispatch which can more efficiently handle the event-free case.
By implementing via extension, DynamicalSystems.jl and its users can
still easily switch to any of the available DifferentialEquations.jl
methods when required. The result is that developers and users have
not had to worry about learning or wrapping many package APIs since
the confederated API is sufficient for calling out to the many choices
in the Julia package ecosystem.

\section{Polyalgorithms for Automatic Algorithm Specialization\label{sec:Polyalgorithms-for-Automatic}}

The confederated formulation of the API has allowed for an explosion
in the available solver methods. As of the writing of this manuscript,
there exist over 200 methods simply for ODEs, and many more when including
the other forms of differential equations offered by DifferentialEquations.jl
(SDEs, DDEs, DAEs, jump equations like Gillespie SSA, etc.). Thus
while ODE solver APIs like those offered by previous packages like
MATLAB \cite{lawrence_f._shampine_matlab_nodate}, SciPy \cite{jones_scipy:_2001},
or R's deSolve \cite{soetaert_package_nodate} all have around 10
methods and document method choosing schemes for users to follow,
we have noticed that our ecosystem's size is far too large to assume
that any user will willingly learn all of the algorithm differences
to make the most effective choice.

\begin{figure}
\begin{centering}
\includegraphics[scale=0.5]{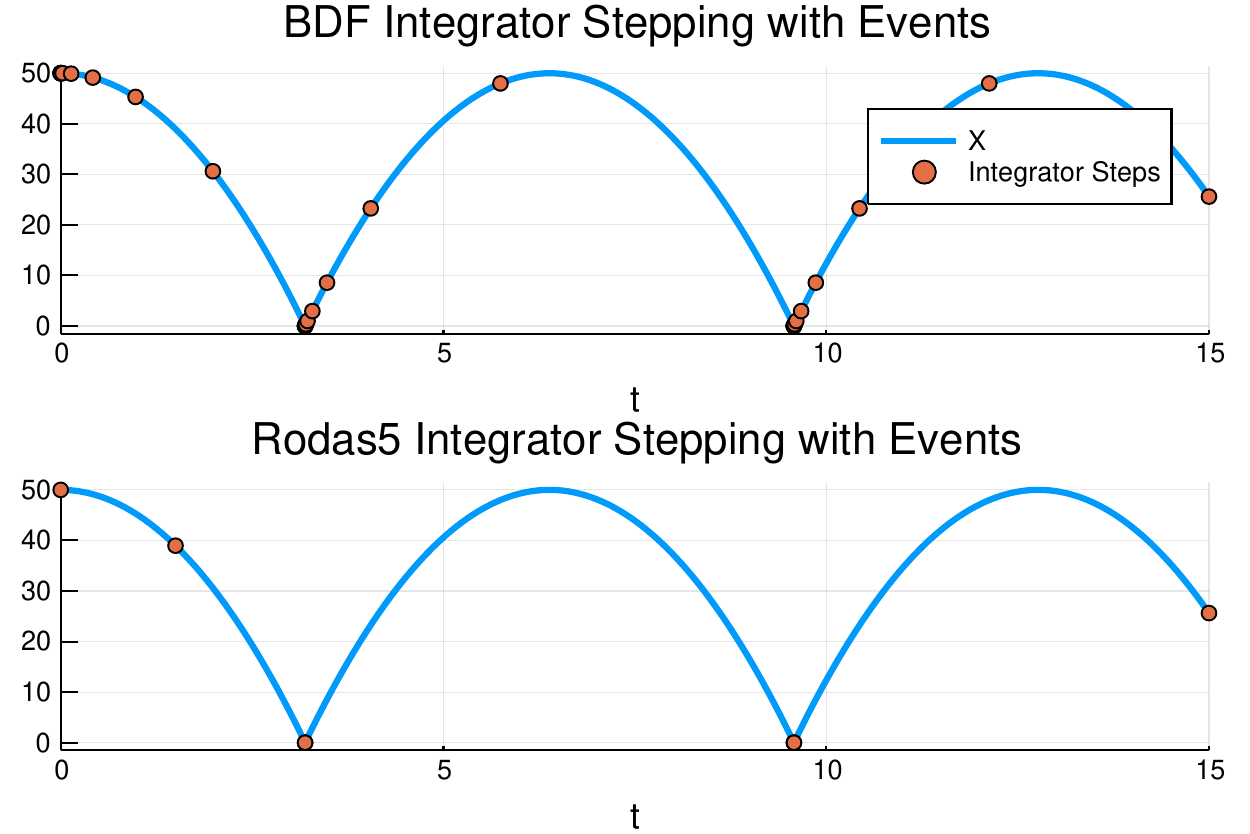}
\par\end{centering}
\caption{\textbf{BDF Integrator Time Stepping with Events. }Shown is the solution
of the bouncing ball hybrid ODE system $v'=-g$, $x'=v$ where $g$
is the gravitational constant $9.8$ and the sign of $v$ flips whenever
$x=0$. In the case of the BDF integrator (SUNDIALS 3.1), events cause
a reset of the order of the integrator to 1, causing 39 time steps
to be taken (shown in orange). Meanwhile, the Rodas5 integrator (OrdinaryDiffEq.jl)
solves the equation in 5 time steps, jumping from event to event and
utilizing an interpolation to fill in the intermediate region. \label{fig:bdf}}
\end{figure}

For example, for large stiff ODE systems a BDF method is recommended
since the multistep method require fewer function evaluations than
methods like ESDIRK. But too large of a system means that a user's
Jacobian will not fit into memory, and thus if no sparse Jacobian
form is given then the user will need to use a low-storage Runge-Kutta
method (or the implicit methods will error due to running out of memory).
Meanwhile, BDF integrators are not as efficient on event-heavy codes
due to the fact that the order must reset to 1 when a derivative discontinuity
is encountered, which results in a time step decrease and increased
computational cost as shown in Figure \ref{fig:bdf}. Thus in this
case one may want to use an ESDIRK method. If the user wants to solve
the equation at low tolerances, the benchmarks frequently show that
the high (and adaptive) order fully implicit Runge-Kutta integrator
Radau \cite{hairer_stiff_1999} is the most efficient, while not being
as efficient at higher tolerances (lower accuracy). Thus, while users
have successfully been walked through these explanations in the JuliaDiffEq
and Julia programming chatrooms, this attention to solver nuances
is not something we have found most users are interested in.

Therefore we have developed polyalgorithms which auto-specialize on
the problem and solver structures to allow for an automatic effective
choice to be made if the user calls the solve function without an
algorithm (i.e. \code{solve(prob)} or with options like \code{solve(prob,reltol=1e-8)}).
Polyalgorithms have been applied in other numerical domains such as
numerical linear algebra where MATLAB and Julia's backslash (\code{\textbackslash{}})
check properties of the matrix before choosing a factorization method.
However, polyalgorithms like this have been non-existent in differential
equation solver software sans a few automated stiffness detection
algorithms which switch between two fixed choices during the time
course of the integration \cite{petzold_automatic_nodate}. Unlike
these previous methods, our algorithm utilizes structural properties
of the solver to choose integration methods, and we also utilize the
user options and allow for hints.

The algorithm starts by dispatching to a choice tree based on the
input problem type, i.e. whether the problem is an ODE, SDE, DAE,
DDE, etc. The dispatch is not only dependent on the type of differential
equation but also its structure. DifferentialEquations.jl for example
allows users to define ODEs in forms such as partitioned ODEs for
second order differential equations and dynamical equations resulting
from Hamiltonian equations, split ODEs ($u'=f_{1}(u,p,t)+f_{2}(u,p,t)$),
and semilinear ODEs ($u'=Au+f(u,p,t)$). These special forms have
an equivalent form as a first order ODE, but this special structure
can be exploited by integrators to give more accurate results, such
as symplectic or Runge-Kutta Nystrom integrators for partitioned ODEs,
implict-explicit (IMEX) integrators for split ODEs, and exponential
integrators for semilinear ODEs. When applicable, the polyalgorithm
will pick an appropriate algorithm from these sub-classes specific
to the problem's structure, or if no structure applies it will fall
back to the first order ODE integrator methods.

\begin{figure}
\begin{centering}
\includegraphics[scale=0.4]{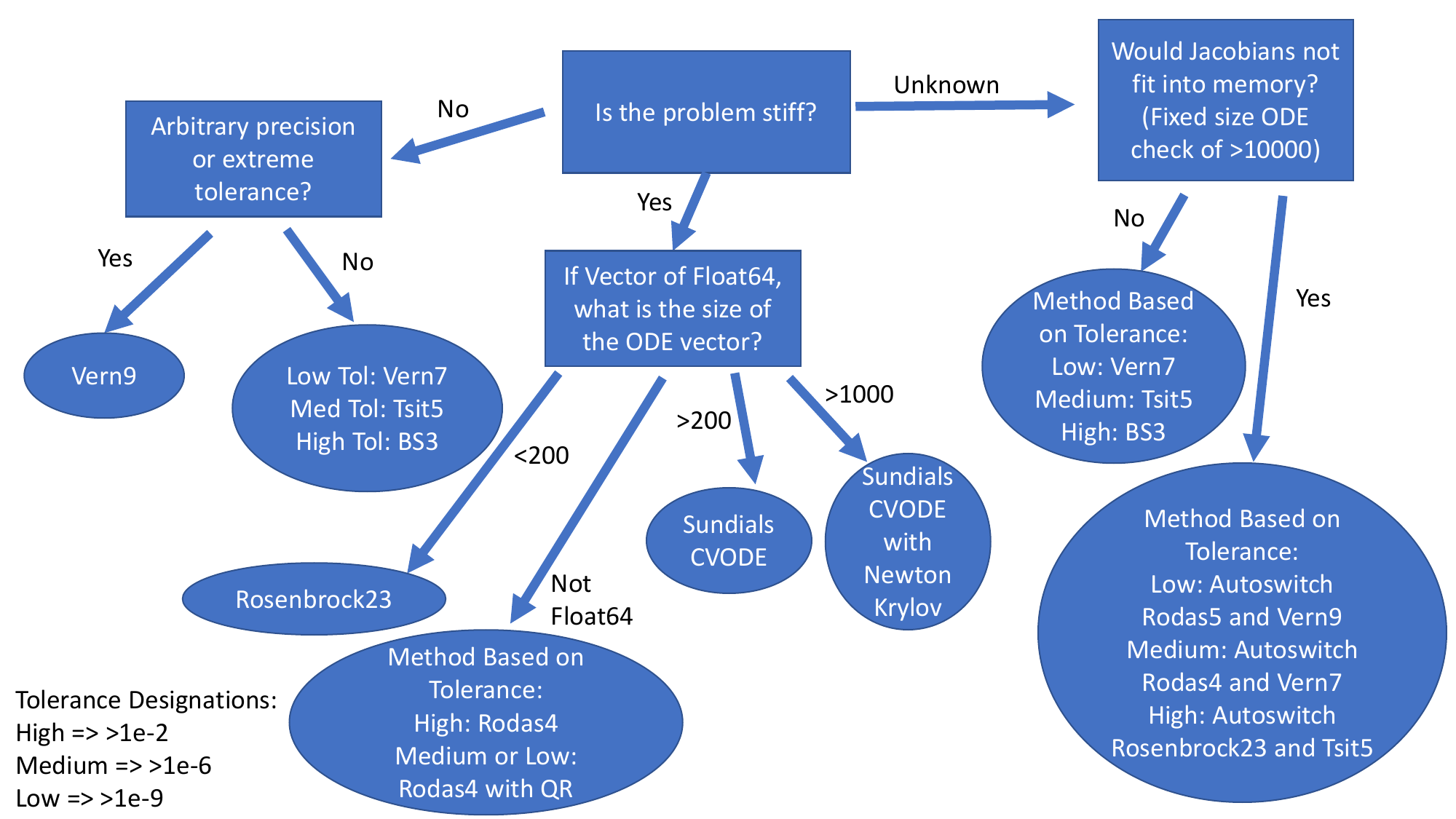}
\par\end{centering}
\caption{\textbf{First order ODE polyalgorithm.} Depicted is the decision tree
for the DifferentialEquation.jl polyalgorithm for choosing the integrator
method based on the user's input tolerances and properties of the
underlying ODE. Note that ``Autoswitch'' denotes an algorithm with
stiffness detection that allows for automated switching between a
method optimized for stiff equations and a method optimized for non-stiff
equations. \label{fig:poly}}

\end{figure}

When choosing a method from a class, features of the problem and solution
are then used to make the final choice. As an example, the current
first order ODE choice tree is shown in Figure \ref{fig:poly} and
makes use of the size of the differential equation system, the user-chosen
tolerance, algorithm hints about stiffness (along with automatic stiffness
detection when not hint is given), and the presence of events. These
choices are heuristics derived from the extensive benchmarks to give
a choice which is likely to be efficient without having to rely on
users having extensive knowledge of the solver details. We have found
that most of the codes from package users utilize the automatic algorithm
choices, whereas code from package developers and methods researchers
(who generally take the time to benchmark methods on their specific
problems) make use of direct algorithm choices. This shows that the
algorithm has been successful in achieving its goal of abstracting
away the numerical nature of differential equation problem solving
while not impeding the flexibility required for developers.

\section{Pervasive Benchmarks for Empirically-Driven Development}

\begin{figure}
\begin{centering}
\includegraphics[scale=0.22]{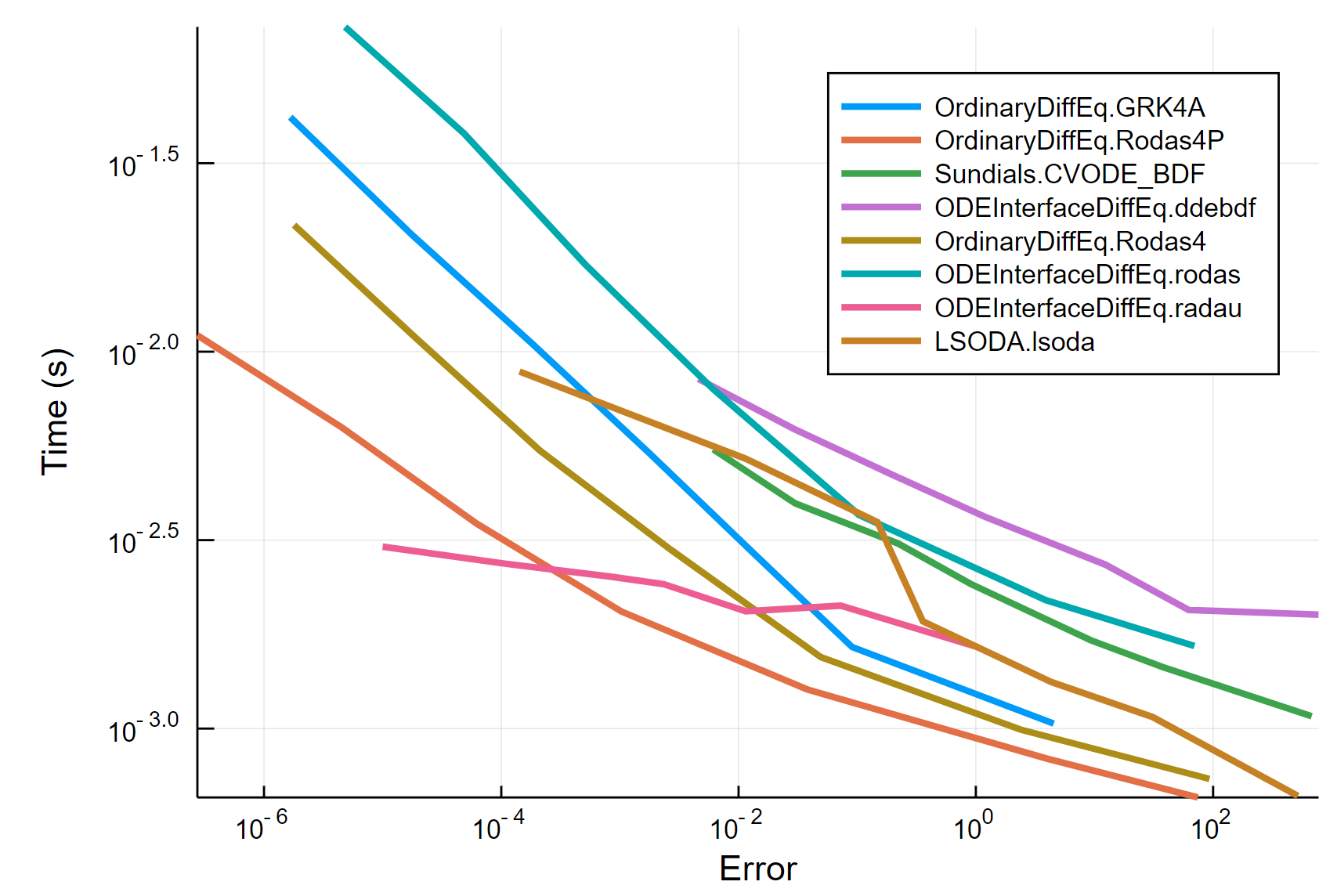}
\par\end{centering}
\caption{\textbf{Example Work-Precision Diagram}. This is a work-precision
diagram of various common interface packages solving the OREGO system
of 3 stiff ODEs \cite{ernst_hairer_solving_1991}. Each algorithm
is documented as \protect\code{Package.algorithm}, for example \protect\code{Sundials.CVODE\_BDF}
refers to the use of CVODE (with BDF coefficients) from the SUNDIALS
C++ package \cite{alan_c._hindmarsh_sundials:_2005}. The error is
the absolute average time series error $\sum_{i}^{n}\Vert\hat{x}(t_{i})-x(t_{i})\Vert/n$
calculated against a reference solution solved with absolute and relative
tolerances at $10^{-14}$. The timings are derived from the benchmark
computer which is a dual Xeon E5-2697A, but as an interactive notebook
users and developers can re-run these benchmarks on their own hardware
to generate comparable figures. \label{fig:Example-Work-Precision-Diagram}}

\end{figure}

Given the wide range and available packages which are directly available
via the common API, we developed a set of integrator analysis tools
in DiffEqDevTools.jl. These tools can directly compute error estimates
by using lower tolerance solutions (and in the case of stochastic
problems, utilize the same Brownian process to measure strong/weak
error), allowing the use of realistic nonlinear models to be used
for benchmarking. A set of benchmarks is shared by the JuliaDiffEq
organization at the JuliaDiffEq/DiffEqBenchmarks.jl repository. An
example work-precision diagram from the stiff ODE benchmark problem
OREGO is shown in Figure \ref{fig:Example-Work-Precision-Diagram}.

We have found that these benchmarks have been useful in guiding the
development of the field. Since a large number of methods and packages
can be pulled from (including classic C++ packages like SUNDIALS \cite{alan_c._hindmarsh_sundials:_2005}
and Fortran codes like Hairer's dop853 \cite{hairer_solving_2009,ernst_hairer_solving_1991}),
these benchmarks are able to effectively isolate differences between
different implementations of the same method and performance enhancements
which are due the integration method itself. While these benchmarks
have been useful to the JuliaDiffEq developers for identifying performance
issues and regressions, these benchmarks have also had a large effect
on our development goals. Observations around the efficiencies of
Rosenbrock and ESDIRK methods for small stiff ODEs and the importance
of coefficient optimization in high order explicit ODE Runge-Kutta
integrators spawned the motivation to optimize stochastic differential
equation integrators of a similar form \cite{rackauckas_stability-optimized_2018}.
Results of benchmarks on LSODA's efficiency vs that of pure BDF methods
was the impetus for a Google Summer of Code project on developing
the automatic stiffness detection and switching in OrdinaryDiffEq.jl.
By having open benchmarks, newcomers to Julia have been able to directly
see how the native Julia implementations fare against those of the
more developed statically compiled languages (C++ and Fortran), which
has spurred other Google Summer of Code students to focus on areas
where JuliaDiffEq was previously lacking such as adaptive multistep
integrators. Additionally, these benchmarks have been helpful in allowing
researchers to assess the necessity of switching their current tooling
by seeing timings on real applications. For example, these benchmarks
routinely show that integrators within the same class (for example,
stiff ODEs looking at BDF vs (E)SDIRK vs Rosenbrock) generally do
not differ in work-precision timing by more than an order of magnitude
with the largest differences being between families of methods rather
than within. Additionally, we see large differences in performance
depending on the language which is used, for example MATLAB ODE's
and R's deSolve suite taking routinely 100x longer, while the SciPy
integrators with Numba-accelerated derivative functions taking routinely
10x longer than the optimized Julia and C++ codes. We hope this verification
and quantification of the Julia performance claims in real-world scenarios
can better help other researchers choose the right path for their
project and team.

From these benchmarks there are many interesting properties about
the families of differential equation solver methods and their performance
on different types of problems which can be elaborated. Our team wishes
to summarize these results in a publication in the near future.

\section{Discussion and Conclusion}

The confederated modular API of the DifferentialEquations.jl package
has had a large impact on the ability for possible methods developers
to contribute to the ecosystem. This in turn has led to the availability
of hundreds of methods from this interface which are directly available
to users of downstream packages and users of the Python and R wrappers.
This catalogue has been and is continuously being benchmarked on real-world
problems in order to guide future JuliaDiffEq development projects
and build an empirically-based polyalgorithm for automatic method
selection.

We see the possibility for these architectures to be employed in other
scientific disciplines for similar effects. While machine learning
packages like SciKitLearn \cite{pedregosa_scikit-learn:_nodate} allow
for calling a large set of methods on the same dataset, a confederated
version built around dispatch could have similar effects on the machine
learning community as we have seen in JuliaDiffEq. The Julia metapackage
JuMP \cite{lubin_computing_2015} for mathematical programming and
optimization has a confederated API, but we believe that it could
have a similar effect on the optimization methods development community
by developing similar cross-method benchmarking tools with continuous
publication of the benchmark results. 

\section{Acknowledgments }

This work was supported by NSF grants DMS 1562176 and DMS1763272 and
a grant from the Simons Foundation (594598, QN).

\bibliographystyle{plain}
\bibliography{references}

\begin{thebibliography}{10}

\bibitem{bezanson_julia:_2017}
J.~Bezanson, A.~Edelman, S.~Karpinski, and V.~Shah.
\newblock Julia: {A} {Fresh} {Approach} to {Numerical} {Computing}.
\newblock {\em SIAM Review}, 59:65--98, January 2017.

\bibitem{biernat_dassl.jl_2016}
Pewel Biernat and Christopher Rackauckas.
\newblock {\em {DASSL}.jl}.
\newblock 2016.

\bibitem{datseris_dynamicalsystems.jl:_2018}
George Datseris.
\newblock {DynamicalSystems}.jl: {A} {Julia} software library for chaos and
  nonlinear dynamics.
\newblock {\em Journal of Open Source Software}, 3(23):598, March 2018.

\bibitem{forst_liblsoda_2016}
Simon Forst.
\newblock {\em liblsoda}.
\newblock 2016.

\bibitem{hairer_solving_2009}
E.~Hairer, S.~P. Norsett, and Gerhard Wanner.
\newblock {\em Solving ordinary differential equations {I}: nonstiff problems}.
\newblock Number~8 in Springer series in computational mathematics. Springer,
  Heidelberg ; London, 2nd rev. ed edition, 2009.
\newblock OCLC: ocn620251790.

\bibitem{ernst_hairer_solving_1991}
Ernst Hairer and Gerhard Wanner.
\newblock {\em Solving {Ordinary} {Differential} {Equations} {II} - {Stiff} and
  {Differential}-{Algebraic} {Problems}}.
\newblock Springer, 1991.

\bibitem{hairer_stiff_1999}
Ernst Hairer and Gerhard Wanner.
\newblock Stiff differential equations solved by {Radau} methods.
\newblock {\em Journal of Computational and Applied Mathematics}, 111:93--111,
  November 1999.

\bibitem{alan_c._hindmarsh_sundials:_2005}
Alan~C. Hindmarsh, Peter~N. Brown, Keith~E. Grant, Steven~L. Lee, Radu Serban,
  Dan~E. Shumaker, and Carol~S. Woodward.
\newblock {SUNDIALS}: {Suite} of nonlinear and differential/algebraic equation
  solvers.
\newblock {\em ACM Trans. Math. Softw.}, 31:363--396, 2005.

\bibitem{jones_scipy:_2001}
Eric Jones, Travis Oliphant, Pearu Peterson, and {others}.
\newblock {\em {SciPy}: {Open} source scientific tools for {Python}}.
\newblock 2001.

\bibitem{kramer_quantumoptics.jl:_2018}
Sebastian Kramer, David Plankensteiner, Laurin Ostermann, and Helmut Ritsch.
\newblock {QuantumOptics}.jl: {A} {Julia} framework for simulating open quantum
  systems.
\newblock {\em Computer Physics Communications}, 227:109--116, June 2018.

\bibitem{lubin_computing_2015}
Miles Lubin and Iain Dunning.
\newblock Computing in {Operations} {Research} {Using} {Julia}.
\newblock {\em INFORMS Journal on Computing}, 27(2):238--248, April 2015.

\bibitem{pedregosa_scikit-learn:_nodate}
Fabian Pedregosa, Gael Varoquaux, Alexandre Gramfort, Vincent Michel, Bertrand
  Thirion, Olivier Grisel, Mathieu Blondel, Peter Prettenhofer, Ron Weiss,
  Vincent Dubourg, Jake Vanderplas, Alexandre Passos, and David Cournapeau.
\newblock Scikit-learn: {Machine} {Learning} in {Python}.
\newblock {\em MACHINE LEARNING IN PYTHON}, page~6.

\bibitem{petzold_automatic_nodate}
Linda Petzold.
\newblock Automatic {Selection} of {Methods} for {Solving} {Stiff} and
  {Nonstiff} {Systems} of {Ordinary} {Differential} {Equations}.

\bibitem{petzold_description_nodate}
Linda~R. Petzold.
\newblock A {Description} of {DASSL}: {A} {Differential}/{Algebraic} {System}
  {Solver}.
\newblock {\em Sandia}, 8637.

\bibitem{christopher_rackauckas_differentialequations.jl_2017}
Christopher Rackauckas and Qing Nie.
\newblock {DifferentialEquations}.jl - {A} {Performant} and {Feature}-{Rich}
  {Ecosystem} for {Solving} {Differential} {Equations} in {Julia}.
\newblock {\em Journal of Open Research Software}, 5:15, 2017.

\bibitem{rackauckas_stability-optimized_2018}
Christopher Rackauckas and Qing Nie.
\newblock Stability-{Optimized} {High} {Order} {Methods} and {Stiffness}
  {Detection} for {Pathwise} {Stiff} {Stochastic} {Differential} {Equations}.
\newblock {\em arXiv:1804.04344 [math]}, April 2018.
\newblock arXiv: 1804.04344.

\bibitem{lawrence_f._shampine_matlab_nodate}
Lawrence~F. Shampine and Mark~W. Reichelt.
\newblock The {MATLAB} {ODE} {Suite}.

\bibitem{soetaert_package_nodate}
Karline Soetaert, Thomas Petzoldt, and R~Woodrow Setzer.
\newblock Package {deSolve}: {Solving} {Initial} {Value} {Differential}
  {Equations} in {R}.
\newblock page~52.

\bibitem{tsitouras_runge-kutta_2011}
Ch. Tsitouras.
\newblock Runge-{Kutta} pairs of order 5(4) satisfying only the first column
  simplifying assumption.
\newblock {\em Computers \& Mathematics with Applications}, 62(2):770--775,
  July 2011.

\bibitem{veltz_lsoda.jl_2017}
Romain Veltz, Christopher Rackauckas, and Simon Frost.
\newblock {\em {LSODA}.jl}.
\newblock 2017.

\end{thebibliography}
\end{document}